\documentclass[jcp,graphicx, reprint]{revtex4-1}
\usepackage{amsmath}
\usepackage{graphicx}
\usepackage{mathtools}

\newcommand{\x}{\boldsymbol{x}}
\newcommand{\m}{\boldsymbol{m}}
\newcommand{\balpha}{\boldsymbol{\alpha}}

\usepackage[usenames,dvipsnames]{xcolor}

\usepackage{comment}

\newif\ifshow
\showtrue
\ifshow

\else
\excludecomment{scratch}
\fi

\draft 

\begin{document}


\title{Bounds on stationary moments in stochastic chemical kinetics} 



\author{Khem Raj Ghusinga}
%
%
\author{Cesar A. Vargas--Garcia}
%

%
\author{Andrew Lamperski}
%
\author{Abhyudai Singh}
\email[]{absingh@udel.edu}


\date{\today}

\begin{abstract}
In the stochastic formulation of chemical kinetics, the stationary moments of the population count of species can be described via a set of linear equations. However, except for some specific cases such as systems with linear reaction propensities, the moment equations are underdetermined as a lower order moment might depend upon a higher order moment. Here, we propose a method to find lower, and upper bounds on stationary moments of molecular counts in a chemical reaction system. The method exploits the fact that statistical moments of any positive-valued random variable must satisfy some constraints. Such constraints can be expressed as nonlinear inequalities on moments in terms of their lower order moments, and solving them in conjugation with the stationary moment equations results in bounds on the moments. Using two examples of biochemical systems, we illustrate that not only one obtains upper and lower bounds on a given stationary moment, but these bounds also improve as one uses more moment equations and utilizes the inequalities for the corresponding higher order moments. Our results provide avenues for development of moment approximations that provide explicit bounds on moment dynamics for systems whose dynamics are otherwise intractable. 
\end{abstract}

\pacs{}

\maketitle 

\section{Introduction} \label{sec:intro}

In a well-mixed chemical reaction system, the reactions are inherently stochastic owing to the perpetual random motion at the molecular level. These stochastic effects are particularly relevant when  the species are present at low molecular counts. Mathematical characterization of such systems is usually done via the Chemical Master Equation (CME) which describes the time evolution of the probability density function for population counts of different chemical species involved \cite{mcq67, kam11, gar85}. However, the CME is analytically intractable for most systems and generally requires significant computational effort if solved via numerical techniques \cite{GillespiePetzoldOct03,RathinamPetzoldCaoGillespieDec03,gil01,gib00,clp04,brk06, and07,hko13}. Often one is interested in estimating the stationary moments such as mean, variance, etc., of different species. For this purpose, one can use the CME to describe the time-evolution of statistical moments of the system via ordinary differential equations, and solve them in steady-state as described below. 

Consider a system of $n$ species $X_j$, $j \in \{1,2,\ldots,n\}$ that interact through $k$ reactions $R_i$, $i\in \{1,2,\ldots,k\}$. We denote the number of molecules of $X_j$ at time $t$ by $x_j(t)$, and use a vector $\x(t)=\begin{bmatrix}x_1(t) & x_2(t) & \ldots & x_n(t) \end{bmatrix}^\top$ to denote the state of the system. Each $R_i$ is assigned a probability $f_i(\x(t))dt$ that it will occur in an infinitesimal small time-interval $[t,t+dt)$. Upon occurrence of $R_i$,  the state of the system is transitioned to $\x-\balpha_i$ where $\balpha_i$ is the stoichiometry vector that describes the change in population as a result of $R_i$. Given a vector $\m=\begin{bmatrix}m_1 & m_2 & \ldots m_n\end{bmatrix}^\top$ of $n$ non-negative integers, a statistical moment of $\x$  is defined as $\left<{x}_1^{m_1}{x}_2^{m_2} \cdots {x}_n^{m_n}\right>$ where the sum $\sum_{j=1}^n{m_{j}}$ is referred to as as the order of the moment. Using a short-hand $\x^{[m]}:={x}_1^{m_1}{x}_2^{m_2} \cdots {x}_n^{m_n}$, one can write the time derivative of $\left<\x^{[\m]}\right>$ obtained from the CME as \cite{hsi04,sih10,sih10a}
\begin{align}\label{dyn}
\frac{d \left< \x^{[\m]}\right>}{dt}& =\left<\sum_{i=1}^k f_i(\x) \left(\left(\x-\balpha_i\right)^{[\m]}-\x^{[m]}\right)\right>.
\end{align}
Assuming the propensity functions  $f_i(\x)$ to be polynomials in elements of $\x$, it follows from  eq.~\eqref{dyn} that if one stacks all statistical moments up to order $M$ in a vector $\mu$, then the time evolution of $\mu$ is given by
\begin{equation}
\frac{d\mu}{dt}=a+A\mu+B\overline{\mu}.
\label{eqn:momdyn}
\end{equation}
Here $\overline{\mu}$ consists of moments of order higher than $M$ \cite{sih10}. The elements of the vector $a$, and the matrices $A$ and $B$ depend upon reaction parameters. As a consequence of eq.~\eqref{eqn:momdyn}, the stationary moments must satisfy the following
\begin{equation}
 a+A\mu+B\overline{\mu}=0.
 \label{eqn:ssmomeq}
 \end{equation}
 
 When the reaction propensities are linear, then $B=0$ and the steady-state moments in $\mu$ can be determined exactly by solving eq.~\eqref{eqn:ssmomeq}. However, in general the matrix $B \neq 0$ which implies that eq.~\eqref{eqn:ssmomeq} represents an underdetermined system of equations. One widely used approach for handling such cases is to employ an appropriate moment closure technique. Based on different assumptions, these techniques approximate the vector $\overline{\mu}$ as a, possibly nonlinear, function of $\mu$ \cite{lkk09,kcm05,sihbmb,sih06cdc,nas03a,nas03,gil09,mgw12,sih06,sih07ny,nfq04,SinghHespanhaJul07,gri12,smk13,svs15,mgw11,lyt07,jdd14,sch_14,sch_15,Lakatos_15,Hase_14,ssg15,bhp15,kue16,lak15}. Although presumed to be reasonably accurate, the moment closure schemes typically do not provide any mathematical guarantees on the accuracy of the approximation.

Here we provide an alternate approach wherein instead of finding an approximation of $\overline{\mu}$, we use some inequalities satisfied its elements. These inequalities are generated from constraints that are required to be satisfied by moments of any random variable. We show that using such inequalities in conjugation with eq.~\eqref{eqn:ssmomeq}, we can find lower and upper bounds on a given moment in $\mu$. Next, we discuss how these inequalities satisfied by the moments can be generated.

\section{Constraints on moments}\label{sec:constraints}
We are interested in determining a bound on a moment of a random variable in terms of lower order moments. For simplicity, let us first consider the case of a scalar random variable $x$. Suppose we construct a vector $v=\begin{bmatrix}1 & x & x^2 & \ldots x^d \end{bmatrix}^\top$ that consists of monomials up to degree $d$ of $x$. Then the outer product $vv^\top$ is positive semidefinite (denoted by $\succeq 0$), and the semidefiniteness is preserved if expectation is taken. That is, we have
\begin{equation}  \label{eqn:vvt} 
\left<vv^\top\right>= 
\begin{bmatrix}1 & \left< x \right> & \ldots & \left< x^d \right> \\ 
\left< x \right> & \left< x^2 \right>  & \ldots & \left< x^{d+1} \right>\\ 
\vdots & \vdots  & \ldots & \vdots \\ 
\left< x^d \right> & \left< x^{d+1} \right> & \ldots & \left< x^{2d} \right> \end{bmatrix} \succeq 0,
\end{equation}
for all $d=\{1, 2, \ldots \}$. Furthermore, if the random variable $x$ is non-negative, another semidefinite constraint can be obtained as
\begin{equation}\label{eqn:xvvt}
\left<xvv^\top \right>= 
\begin{bmatrix}\left<x\right> & \left< x^2 \right> & \ldots & \left< x^{d+1} \right> \\ 
\left< x^2 \right> & \left< x^3 \right>  & \ldots & \left< x^{d+2} \right>\\ 
\vdots & \vdots  & \ldots & \vdots \\ 
\left< x^{d+1} \right> & \left< x^{d+2} \right> & \ldots & \left< x^{2d+1} \right> \end{bmatrix} \succeq 0,
\end{equation}
for all $d=\{1, 2, \ldots \}$. Note that a matrix is positive semidefinite iff all its leading principal minors are non-negative. In the case of $d=1$, the non-negative determinants of the matrices in eq.~\eqref{eqn:vvt} and eq.~\eqref{eqn:xvvt} result in   
\begin{equation}
\left<x^2\right>\geq \left<x\right>^2, \quad \left<x^3\right>\geq \frac{\left<x^2\right>^2}{\left<x\right>}.
\label{eqn:1dineq}
\end{equation}
respectively. Note that the first inequality above is nothing but the well-known inequality representing non-negativity of variance. Similarly, for $d=2$, the determinant of the matrix in eq.~\eqref{eqn:vvt} yields
\begin{equation} 
\left<x^4\right> \geq \frac{\left<x^3\right>^2 + \left<x^2\right>^3 - 2\left<x^3\right>\left<x^2\right>\left<x\right>}{\left<x^2\right>-\left<x\right>^2}.
\end{equation}
In essence, these determinants for varying $d$ allow higher-order moments to be bounded from below by nonlinear functions of the lower-order moments. Another point to note is that the matrix $\left<vv^\top\right>$ generates inequalities for even order moments whereas $\left<xvv^\top\right>$ generates inequalities for odd order moments. In principle, one could also generate constraints from positive semidefiniteness of $\left<g(x)vv^\top\right>$, where $g$ is a non-negative polynomial function of $x$.

Next, we consider a $n$-dimensional multivariate random variable $\x=\begin{bmatrix}x_1 & x_2 & \ldots & x_n \end{bmatrix}^\top$, and discuss how we can write inequalities that bound moments of the form $\left<{x}_1^{m_1}{x}_2^{m_2} \cdots {x}_n^{m_n}\right>$. To this end, we note that a matrix analogous to the one in eq.~\eqref{eqn:vvt} can be constructed by taking expectation of outer product $vv^\top$ where the vector $v$ consists of all monomials of $\x$ up to order $d$
\begin{equation}
\small
v=\begin{bmatrix}
1 & x_1  \ldots  x_n & x_1^2 & x_1x_2  \ldots  x_1x_n  \ldots  x_n^2   \ldots  x_n^d
\end{bmatrix}^\top.
\label{eqn:v}
\end{equation}
In this case, non-negativity of the principal minors of matrix $\left<vv^\top\right>$ gives bound on the moments $\left<x_i^{2d}\right>$, $i= \{1,2,\ldots,n \}$. As an example, for $\x= \begin{bmatrix}x_1 & x_2 \end{bmatrix}^\top$, the following is obtained for $d=1$
\begin{align}
\left<\begin{bmatrix}1 & x_1 & x_2  \\ x_1 & x_1^2 & x_1 x_2 \\ x_2 & x_1 x_2 & x_2^2  \end{bmatrix}\right> \succeq 0,
\end{align} 
which results in $\left<x_1^2\right>\geq \left<x_1\right>^2$, and the following inequality bounding $\left<x_2^2\right>$
\begin{equation}
\left<x_2^2\right> \geq \frac{\left<x_1x_2\right>^2+\left<x_1^2\right>\left<x_2\right>-2\left<x_1\right>\left<x_2\right>\left<x_1x_2\right>}{\left<x_1^2\right>-\left<x_1\right>^2}.
\end{equation}

To get a bound on moments whose form is different from $\left<x_i^{2d}\right>$, we can take expectation of $\left<g(\x)vv^\top\right>$ where $g$ is an appropriate positive valued polynomial function  \cite{las09}. For instance, when $\x$ takes positive values, $g(\x)=x_1$ gives the constraint
\begin{align}
\left<x_1 \begin{bmatrix}1 & x_1 & x_2 \\ x_1 & x_1^2 & x_1 x_2  \\ x_2 & x_1 x_2 & x_2^2 \end{bmatrix}\right> \succeq 0.
\end{align}
that can be used to find a lower bound on $\left<x_1x_2^2\right>$.

It should also be noted that the univariate inequalities obtained from eqs.~\eqref{eqn:vvt}--\eqref{eqn:xvvt} are valid for both $x_1$ and $x_2$. Furthermore, as $x_1$ and $x_2$ are positive random variables, additional inequalities as follows can also be written
\begin{equation}
\left<x_1 \begin{bmatrix}1 & x_2 \\ x_2 & x_2^2 \end{bmatrix}\right> \succeq 0, \quad \left<x_2 \begin{bmatrix}1 & x_1 \\ x_1 & x_1^2 \end{bmatrix}\right> \succeq 0,
\label{eqn:mommatex2}
\end{equation}
These constraints can also be used to construct lower bounds on moments of the form $\left<x_1^2x_2\right>$, and $\left<x_1x_2^2\right>$.
To sum up, one can write a multitude of constraints satisfied by the moments using moment matrices. In the following section, we discuss how these inequalities can be used to find bounds on moments.

\section{Bounds on Steady-State Moments}\label{sec:bounds}
In this section, we use two examples to demonstrate how bounds on a certain moment can be obtained by using the inequalities obtained from the moment matrices in conjugation with eq.~\eqref{eqn:ssmomeq}.

\subsection{Example 1: Stochastic logistic growth with constant immigration rate}
Consider the following reactions, where a species $X$ is synthesized at a rate $\alpha+rx$, and it degrades with a nonlinear rate $\frac{r}{C}x^2$
   \begin{equation}\label{eqn:logistic}
    X \xrightarrow{~\alpha+rx~} {X+1},  \ \ X\xrightarrow{~rx^2/C~} {X-1},
    \end{equation}
where $x$ denotes the population level of the species. With $\alpha=0$, this model essentially represents a logistic growth model which is widely used to model growth of populations in ecology, and virus dynamics \cite{All03,mak99,sihbmb,nas01}. In the deterministic sense, the population grows with a rate $r$ and saturates once it reaches a finite carrying capacity $C$ due to resource limitations. The term $\alpha$ here represents a constant rate of immigration so as to avoid the extinction of the population. Our aim is to find bounds on the steady-state mean $\left<x\right>$.

\begin{figure}
\centering
\includegraphics[width=\linewidth]{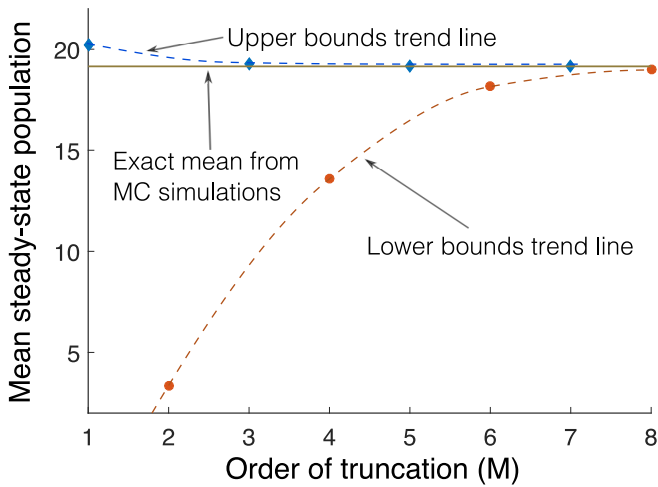}
\caption{ {\it Estimated upper and lower bounds of the steady-state mean population level}.  The upper bounds (diamonds), and the lower bounds (circles) on the actual population level are shown for different orders of truncation $M$. The bounds for $M=1$, and $M=2$, respectively, come from \eqref{eqn:ub}, and \eqref{eqn:lb}. The exact mean value of $19.149$ (standard error of mean$=4.494\times 10^{-5}$) is obtained by averaging $100,000$ MC simulations performed using SSA \cite{Gillespie76}. As $M$ is increased, the bounds obtained get tighter as highlighted by the dashed trend lines. Parameters in \eqref{eqn:logistic} taken as $\alpha=1$, $r=5$, and $C=20$ with an of initial population count  of zero with probability one.}
\label{fig:logistic}
\end{figure}

At steady-state, the time derivatives of the first two moments satisfy
\begin{align}
\dot{\langle x \rangle}&=\alpha + r \left<x\right> - \frac{r}{C} \left<x^2\right> =0, \\
\dot{\langle x^2 \rangle}&=\alpha +\left(2\alpha +\frac{r}{C}\right) \left< x \right>+\left(2r+\frac{r}{C}\right) \left< x^2 \right> \nonumber \\
                                     & \qquad \qquad \qquad \qquad \qquad  -\frac{2r}{C} \left<x^{3}\right> =0.
\end{align} 
Notice that these equations are not closed because of the quadratic propensity function. To obtain a bound on $\left<x\right>$, we solve the expression $\dot{\left<x\right>}=0$ which gives $\left<x^{2}\right>=C\left(\alpha + r \left<x\right>\right)/r $. Using this with the non-negative variance inequality $\left<x^2\right> \geq \left<x\right>^2$ yields
\begin{equation}  \label{eqn:ub}
\langle x \rangle \leq \frac{1}{2} \sqrt{\frac{4 \alpha C+C^2 r}{r}}+\frac{C}{2},
\end{equation} 
which is an upper bound to the actual steady-state mean (value corresponding to $M=1$ in Fig.~\ref{fig:logistic}). Similarly, solving $\dot{\left<x\right>}=0$ and $\dot{\left<x^2\right>}=0$ results in 
\begin{equation}\label{eqn:xx2}
\small
\left<x\right>=-\frac{\alpha C^2+\alpha C-r \left<x^3\right>}{C (\alpha+C r+r)}, \left<x^2\right>=\frac{\alpha^2 C+r^2 \left<x^3\right>}{r (\alpha+Cr+r)}.
\end{equation}
Using the first inequality from \eqref{eqn:1dineq} in this expression leads to a quadratic inequality in $\left<x\right>$, one solution of which gives an lower bound on $\left<x\right>$ (corresponding to M=2 in Fig.~\ref{fig:logistic}) as below
\begin{equation}  \label{eqn:lb}
\langle x \rangle \geq \frac{1}{2} \sqrt{\frac{4 \alpha^3 C+\alpha^2 C^2 r+2 \alpha^2 C r+\alpha^2 r}{r (\alpha+r)^2}}+\frac{\alpha (C-1)}{2 (\alpha+r)},
\end{equation} 
while the other solution is discarded using the fact that $\left<x\right>\geq 0$. One can also find a bound on $\left<x^2\right>$ by using the relations from \eqref{eqn:xx2} to find $\left<x^2\right>$ in terms of $\left<x\right>$ and then using the bounds from \eqref{eqn:lb}. Thus, determining bound on $\left<x\right>$ suffices to find bounds on other lower order moments.

In the same fashion as above, we can consider steady-state equations of first $M$ moments and use the inequality bounding the $(M+1)^{th}$ moment. Though the resulting expressions do not lead to closed-form analytical bounds like \eqref{eqn:ub} and \eqref{eqn:lb}, numerical solutions are still possible. Interestingly, the solutions for odd (even) values of $M=3,5,7$ ($M=4,6,8$) provide a decreasing (an increasing) sequence of upper (lower) bounds on the average population level (Fig.~\ref{fig:logistic}). The lower and upper bounds using $8^{th}$  and $7^{th}$ order truncations are respectively given by $18.9711$ and $19.1659$. The exact average population level obtained from Monte Carlo simulations of the process is $19.1495$ with standard error of mean $4.4940\times 10^{-5}$. As discussed in context of $M=2$, the bounds on $\left<x\right>$ automatically translate to bounds on other lower order moments up to order $M$.

\subsection{Example 2: Stochastic gene expression with negative auto-regulation}
Consider a stochastic expression of an auto-regulating gene that can be represented by the following reactions
\begin{equation}
{Gene_{OFF}} \xrightleftharpoons[k_{off}x_2]{\,k_{on}\,} {Gene_{ON}},  Gene_{ON} \xrightarrow{~\gamma_p~} {B \times Protein}.
\label{eqn:geneexp}
\end{equation}
Here a gene is assumed to reside in one of the two states: ON (active), and OFF (inactive). The protein is transcribed in geometric bursts (denoted by $B$) from the ON state whereas there is no protein production when the gene is in the OFF state. The gene state is represented by $x_1$ which is a Bernoulli random variable ($x_1=1(0)$ for the ON (OFF) state) while the protein level is represented by $x_2$. The gene negatively regulates itself by switching to OFF state in a protein copy number dependent fashion with a rate $k_{off}x_2$. Such gene expression models are known to be experimentally verified and are widely studied in the literature \cite{sih09c,svn14,kpk14,svs15}. 

\begin{figure}
\centering
\includegraphics[width=\linewidth]{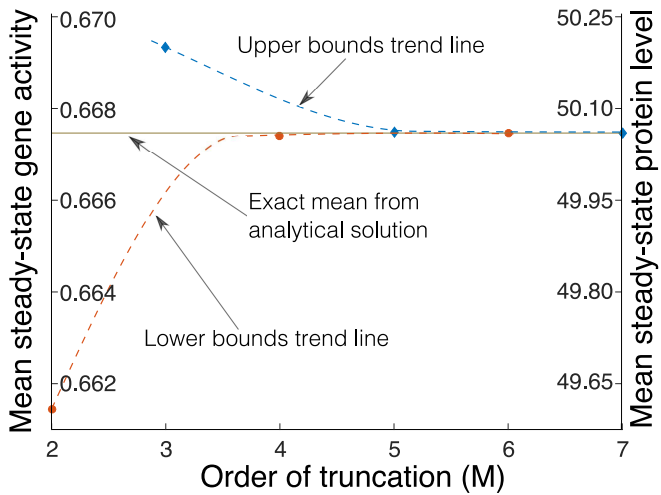}
\caption{ {\it Estimated upper and lower bounds of the steady-state gene activity and mean protein level for a stochastic gene expression model}. The upper bounds (diamonds), and the lower bounds (circles) on the actual gene activity/protein level are shown for different orders of truncation $M$. The exact mean values are obtained from the exact analytical solution of the system from \cite{kpk14}. As $M$ is increased, the bounds obtained get tighter as highlighted by the dashed trend lines. Parameters in eq.~\ref{eqn:geneexp} taken as $k_{on}=1$, $k_{off}=4$, $k_p=10$, $\left<B\right>=1$, and $\gamma_p=1$ with an of initial population count  of zero molecules with probability one.}
\label{fig:geneexp}
\end{figure}

As with the one-dimensional example, here too we are interested in obtaining bounds on the mean of the state $\x=[x_1, x_2]^\top$. Towards this end, we begin by writing moment first order moment equations in steady-state
\begin{align}
\dot{\left<x_1\right>}&=k_{on}-k_{on}\left<x_1\right>-k_{off} \left<x_1 x_2\right>=0,\\
\dot{\left<x_2\right>}&=k_p \left<B\right> \left<x_1\right> - \gamma_p \left<x_2\right>=0.
\end{align}
Here the moment equations are not closed due to nonlinearity arising because of the negative feedback. To obtain a bound on $\left<x_1\right>$ and $\left<x_2\right>$, we require a bound on the second order moment $\left<x_1 x_2\right>$. Note that $x_1 \in \{0,1\}$ is a binary random variable, it results in following relations
\begin{align}
\left<x_1^i\right>&=\left<x_1\right>, i \in \{1,2,3,\ldots\},\\
\left<x_1^i x_2^j\right>&=\left<x_1 x_2^j\right>, i,j \in \{1,2,3,\ldots\}.
\end{align}
Thus, using $\left<x_1^2x_2\right>=\left<x_1x_2\right>$ and inequality obtained from first matrix of Eq.~\eqref{eqn:mommatex2}, a bound $\left<x_1x_2\right> \leq \left<x_2\right>$ can be found. Plugging this in the moment equations yields
\begin{equation}
\left<x_1\right> \geq \frac{k_{on} \gamma_p}{k_{on} \gamma_p + k_{off} k_p \left<B\right>}.
\label{eqn:bound1}
\end{equation}

In the similar fashion as above, we can write moment equations up to order two and use the inequality for $\left<x_1 x_2^2\right>$ obtained from the second matrix of Eq.~\eqref{eqn:mommatex2}. This eventually leads to another set of lower bounds on both $\left<x_1\right>$ and $\left<x_2\right>$ ($M=2$ in Fig. 2).
As expected this bounds shows significant improvements from that obtained using the first-order moment equations in Eq.~\eqref{eqn:bound1}. Continuing in similar way, we obtain an increasing sequence of lower bounds, and a decreasing sequence of upper bounds as the order of moment equations $M$ is increased (Fig. 2). The bounds obtained for $\left<x_1\right>$ via $6^{th}$ and $7^{th}$ order truncation are $0.667463$ and $0.667465$ respectively. These are extremely precise as the exact solution for $\left<x_1\right>$ is $0.667464$ as obtained using the exact solution from \cite{kpk14}.

\section{Discussion}
In this paper, we propose a method to obtain both lower and upper bounds on stationary moments of a chemical reaction system. The method uses the steady-state moment equations obtained from the chemical master equation along with inequalities that are required to be satisfied by moments of a random variable. These inequalities are constructed from positive semidefinite constraints on moments of a positive random variable. Using two examples of biochemical reaction systems, we show that not only one can obtain upper and lower bounds on a given stationary moment, but also both upper and lower bounds improve considerably as one uses more moment equations. Intuitively the improvement in the bounds by increasing the order of truncation $M$ stems from two components: including more moment equations gives more information about the underlying chemical reaction system, and including more inequalities puts more constrains on the values that elements of $\overline{\mu}$ can take.

The examples also illustrate that whether using a certain inequality will yield a lower bound or an upper bound depends upon the structure of the problem. For example, in the one-dimensional example of stochastic logistic growth, the truncations at $M=1$ and $M=2$ respectively yield upper and lower bounds. On the other hand, in the two-dimensional example of gene expression with feedback regulation, the truncations at $M=1$ and $M=2$ both result in lower bounds. Future work will systematically address the effect of structures of the vector $a$, and the matrices $A$ and $B$ on the nature of bound obtained. We will further extend this approach to obtain bounds on dynamics of moments of a stochastic process. 


%
%
\begin{acknowledgments}
The authors would like to thank Mohammad Soltani for discussions on some of the ideas contained herein.
\end{acknowledgments}

\bibliography{RefMaster}

\end{document}